\documentclass[onecolumn]{elsart}

\usepackage{natbib}

\usepackage{graphicx}

 \usepackage{amssymb}
 \usepackage{amsmath}

\usepackage{array}

\begin{document}

\begin{frontmatter}



\title{Tidal pattern instabilities on  multi-moon planets}


\author[wlc]{Joanna Furno}
\author[wlc]{Kerry Kuehn}
\address[wlc]{Department of Physical Sciences, Wisconsin Lutheran College, Milwaukee, WI 53226 (U.S.A.)}
\ead{kerry.kuehn@wlc.edu}
\ead[url]{http://faculty.wlc.edu/kuehn/index.html}

\begin{center}
\scriptsize
\end{center}


\begin{abstract}
The equilibrium tide-generating forces in the lunar orbital plane of a planet of radius $R$ are calculated for the case of $N$ moons of mass $M_i$ orbiting the planet at instantaneous polar coordinates $(D_i$, $\alpha_i)$.  For the case of a single moon, there are only two high tides.  For the case of two moons, it is found that there can exist a critical lunar orbital distance at which two high tides become unstable with respect to formation of three high tides.  Bifurcation diagrams are presented which depict how the angular positions of the high and low tides on the planet vary with the lunar distances and lunar separation angle.  Tidal stability diagrams, which illustrate the stability regions for various tidal patterns as a function of lunar distances and lunar separation angle, are presented for various values of $D_2/D_1$ and $M_2/M_1$.  Generally speaking, the aforementioned tidal instability, and hence the propensity for formation of three high tides on a two-moon planet, exists over a significant range of lunar distances and separation angles provided that $M_2/M_1\sim(D_2/D_1)^3$. For the case of $N>2$ moons, the tidal stability diagram becomes more complex, revealing a diversity of potential tidal patterns.
\end{abstract}
 


\end{frontmatter}

\clearpage

\section{Introduction}
\label{sec:intro}

High precision measurements have led to the detection of over 170 extrasolar planets in recent years \citep{Perryman:2000fk,Butler:2006mi}.  Theoretical work has been devoted to developing models of planetary formation \citep{Wetherill:1996tw, Ida:2004zr, Ida:2005lh, Raymond:2006kl, Yamoto:2006qa, Kokubo:2006ff}, and to assessing the likelihood that terrestrial extra-solar planets have liquid on their surface and are habitable \citep{Kasting:1993pi, Gaidos:2004uq, Ikoma:2006uq}.  Moreover, numerical simulations have demonstrated that a giant-impact scenario \citep{Hartmann:1975fk, Cameron:1976uq} can, in certain cases, give rise to multiple moons orbiting such a terrestrial planet \citep{Ida:1997fk, Canup:1999fj}.  Regardless of the causal history, it is manifestly clear that terrestrial planets can in fact be orbited by multiple moons.  For example, Deimos and Phobos orbit Mars.  If such a terrestrial planet were covered by liquid water, one would expect to observe ocean tides on the surface of the planet.

In this paper, we investigate, from a theoretical viewpoint, the tidal pattern on putative multi-moon planets covered with a uniform layer of fluid.  By tidal pattern we mean the total number of high and low tides which result from the azimuthal component of the tide-generating forces.   Our results suggest that unexpected tidal instabilities arise due to gravitational gradients produced by multiple moons in the vicinity of the planet.   In particular, there typically exist one or more critical lunar distances at which bifurcations occur between situations in which different numbers of high tides are stable.  To our knowledge, these instabilities have not hitherto been systematically investigated.

\section{Background}
\label{sec:bg}

Tidal patterns in an ocean covering a planet can be very complex.   These complexities arise due to local meteorological and geographic conditions, multiple nearby gravitating bodies executing non-trivial orbital patterns around the planet, and fluid inertia, friction, and viscosity (for a review, see \citet*{Cartwright:1977kx}).  In the absence of all of these complicating factors, it is believed that the tidal patterns themselves would be much simpler.  \citet{Newton:1952wm} demonstrated that a single gravitating body located near a planet will induce two tidal bulges, one on either side of the planet.  Thus, his theory predicts semi-diurnal tides.  Moreover, the presence of a second gravitating body in the neighborhood of the planet can enhance the tidal bulges when the second body is in conjunction (or opposition) with the first, or diminish the tidal bulges when it is in quadrature with the first.  Thus, his theory also predicts spring and neap tides.  Newton's was an equilibrium theory, in that he neglected the inertia of the water, assuming the height of the tides to be completely determined by the instantaneous positions of the nearby gravitating bodies.

An important failure of Newton's theory was its inability to explain the general lack of temporal coincidence of high tides at a particular meridian with the instant at which the moon is directly overhead. \citet{Laplace:1969kx} attempted to describe the dynamic response of the oceans to a simple periodic tide-generating force by developing a non-equilibrium theory of the ocean tides.  The real tide-generating force, however, is not simple because, among other things, the moon and the sun themselves exhibit non-trivial orbits relative to the earth.  In order to deal with complex tidal forces, the method of harmonic analysis was developed \citep{Kelvin:1910fk, Darwin:1962fk, Doodson:1921fk, Munk:1966fk}.  This technique consists of reconstructing a tide-generating force from the spectral components of an observed tidal time-series at a particular geographical location.  To each spectral component, or partial wave, one attempts to identify a particular periodic cause, such as the daily rotation of the earth, the monthly orbit of the moon, or the creep of the moon's nodes along the solar ecliptic.  

We would like to emphasize that from the viewpoint of harmonic analysis, the complexity of the tidal pattern is thought to arise as a result of, first, the complexity of the relative motion of the earth, the moon, and the sun, and second, the dynamic response of the water to the tide generating forces.  On the other hand, Newton's equilibrium theory is thought to predict relatively simple tidal phenomena, namely the existence of semi-diurnal tides, spring tides, and neap tides.  In this paper we demonstrate that, even within the framework of Newton's equilibrium theory, non-trivial tidal patterns can arise on the surface of a fluid covered planet due to the action of multiple gravitating bodies in its vicinity.  

The paper is organized as follows.  In Sec.~\ref{sec:forces}, we first enumerate the assumptions of our equilibrium planetary model.  We then derive the tide-generating potential function from first principles by considering the forces acting on a point on the surface of a planet orbited by a single moon at distance $D$.  We perform this derivations so as to clearly illustrate that, within equilibrium theory, the tide-generating forces caused by a single moon can not give rise to more than two high tides in the lunar orbital plane of the planet.  We then generalize our equations so as to account for the tide-generating forces caused by $N$ moons.  

In Sec.~\ref{sec:2moons}, we study the case of two moons orbiting a planet of radius $R$.  First, we consider the special case of two identical moons orbiting the planet at identical distances, $D$, and at angular positions $+\alpha$ and $-\alpha$.  We demonstrate the appearance of a tidal instability at a particular value of $R/D$, which depends upon the lunar separation half-angle, $\alpha$.  This instability separates regions of parameter space in which the fluid surface is characterized by different numbers of high tides.  Physically, the angular separation of the moons and their proximity to the planet's surface dictate whether the higher order harmonics in the tide-generating potential are able to induce tidal separation.  By systematically varying these parameters, we can calculate the number and angular positions of the high and low tides for different values of $R/D$ and $\alpha$.   The results of these calculations are illustrated in a number of bifurcation diagrams.  

We then relax the assumption that the moons orbit at identical distances, so as to study whether the tidal instability persists for more realistic scenarios.  It does, and the results of this investigation are presented in a stability diagram.  The stability diagram depicts the location of the transition, in parameter space, from two to three high tides, as a function of the lunar separation half-angle.  If the moons are at different orbital distances, they will of course be orbiting at different rates, and hence the lunar separation angle will be gradually changing.  But within equilibrium theory, the tides will adopt a pattern determined by the instantaneous position of the moons.  

Next, we consider the case in which both the orbital distances and the lunar masses differ.  We find again that the instability persists.   We also derive a law governing the zero stability point, which is defined in the text.  Next, we check our results to make sure that they do not contradict the tidal behavior observed on earth due to the combined action of the moon and the sun.  We also speculate about the types of planetary systems which might demonstrate tidal instabilities such as are predicted in this paper.  

In Sec.~\ref{sec:3moons}, we briefly consider the case of three moons orbiting a planet at identical distances, and we present the associated stability diagram.  Finally, in Sec.~\ref{sec:conclusion}, we summarize our results and highlight the differences between ours and those of previous theoretical studies of tide formation.

\section{Tide generating potential}
\label{sec:forces} 

\begin{figure}[tbp]
\begin{center}
\includegraphics[angle=0,width =3in]{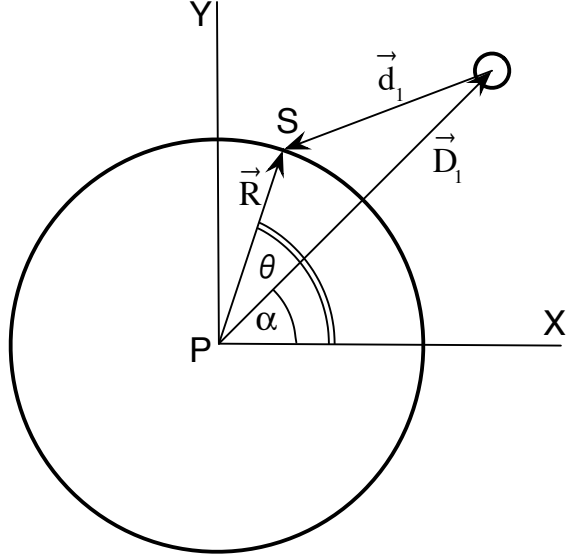}
\caption[fig1]{
\label{fig:fig1}
Schematic diagram of a single moon orbiting a planet.  The origin of the cartesian coordinate system, $XY$, is placed at point $P$, the center of the planet.  Vector $\vec{R}$ indicates a point, $S$, on the surface of the planet.  The vector $\vec{D}_1$ indicates the instantaneous position of the moon.  The vector $\vec{d}_1$ is the difference between $\vec{R}$ and $\vec{D}_1$.  Positive angles are measured clockwise from the abscissa.
}
\end{center}
\end{figure}

Consider the situation in which a single moon orbits a planet, as depicted in Fig.~\ref{fig:fig1}.  We assume that the planet is perfectly rigid and that its mass distribution is spherically symmetric about the center of mass, $P$.  We also assume, as did Newton, that the tidal pattern is determined by the instantaneous positions of the local masses.  The masses of the planet and the moon are $M_P$ and $M_1$, respectively.  A cartesian coordinate system, $XY$, is fixed  at the center of the planet.  This coordinate system lies in the plane of the lunar orbit.  For simplicity, we take the lunar orbital plane to be coincident with the equatorial plane of the planet.  The vector $\vec{D}_1$ indicates the instantaneous position of the orbiting moon.  The coordinate system is oriented such that this vector makes an angle $\alpha_1$ with the abscissa.  Positive angles are measured clockwise from the abscissa.  The vector $\vec{R}$ points to an arbitrary point, $S$, on the surface of the planet, and makes an angle, $\theta$, with the abscissa.  The vector $\vec{d}_1$ points from the moon to the point $S$.  For simplicity, we consider all vectors to lie in the lunar orbital plane.  Thus our analysis is confined to the tidal pattern which arises on the planetary equator.\footnote{ In the event that the orbital and equatorial planes are obliquely oriented, the computed tidal pattern will be in the orbital, not the equatorial, plane.  Even within the context of equilibrium theory, the tides will be distorted by the dependence of the centrifugal force on longitude, but the number of low and high tides will not be affected, since the centrifugal force is purely radial, and it is the azimuthal force that causes tidal separation.}

A unit mass at point $S$ experiences forces due to the gravitational attraction of the planet, $\vec{F}_P$, and the moon, $\vec{F}_1$.
\begin{eqnarray}\label{eq:f1}
 \vec{F}_P & = & - GM_P\vec{R}/R^3  \nonumber \\
 \vec{F}_1 & = & -GM_1\vec{d}_1/d_1^3
\end{eqnarray}
Here, $G$ is the universal gravitational contant.  There also arise fictitious forces on the unit mass due to the rotation of the planet, $\vec{F}_{\omega}$, and due to the fact that the planet is falling toward the moon, $\vec{F}_1'$.
\begin{eqnarray}\label{eq:f2}
\vec{F}_{\omega} & = & \omega^2 \vec{R} \nonumber \\
\vec{F}_1' & = & GM_1\vec{D}_1/D_1^3
\end{eqnarray}
In $\vec{F}_{\omega}$, $\omega$ describes the angular speed of the planet's rotation.  This term, which represents the centrifugal force exerted at point $S$, is purely radially directed, and hence cannot induce tidal separation.  Adding the above equations, and neglecting the manifestly radially directed forces, we find that the net tide-generating force is just the sum of the attractive, gravitational forces and the repulsive, non-inertial forces:
\begin{equation}\label{eq:f3}
	 \vec{F}  =   G M_1 \left( \vec{D}_1/D_1^3 - \vec{d}_1/d_1^3  \right)
\end{equation}
Eq.~\ref{eq:f3} still contains both azimuthal and radial components; we wish to isolate the azimuthal component.  We do this by first writing all vectors in terms of their cartesian components.  Using vector addition, we write
\begin{equation}
\vec{d}_{1}   =  \left[R\cos{(\theta}) - D_1\cos{(\alpha_1})\right]\hat{x}  +  \left[R\sin{(\theta}) - D_1\sin{(\alpha_1})\right]\hat{y}. \nonumber
\end{equation}
Also, the magnitude of $\vec{d}_1$ can be written in terms of $R$, $D_1$, $\alpha_1$ and $\theta$.
\begin{equation}
{d_1}^2   =  {D_1}^2 + R^2 - 2D_1 R\cos{(\theta - \alpha_1)} 
\end{equation}
Substituting these into Eq.~\ref{eq:f3}, and projecting it onto a unit vector tangent to the planet's surface yields
\begin{equation}\label{eq:f4}
	F_{\theta}  =  G M_1\left( \frac{\sin{(\theta - \alpha_1)}}{{D_1}^2}    - \frac{D_1\sin{(\theta - \alpha_1)}}{[{D_1}^2+R^2-2D_1 R\cos{(\theta-\alpha_1)}]^{3/2}}  \right).
\end{equation}
Eq.~\ref{eq:f4} describes the tide-generating forces acting in the lunar orbital plane of a planet orbited by a single moon of mass $M_1$ with instantaneous polar coordinates $(D_1, \alpha_1)$.  If there is a thin layer of fluid covering the surface of the planet at radius $R$, this fluid can experience tidal separation due to these forces.  They tend to heap up (or spread out) water at particular angular locations, causing high tides (or low tides) to form.  

In the literature, it is common to infer the tide-generating forces from a tide-generating potential function.  From our perspective, the tide-generating potential function can be obtained by integrating Eq.~\ref{eq:f4} with respect $\theta$:
\begin{eqnarray}\label{eq:potfunc}
	V  & = &  -\frac{G M_1}{D_1} \times  \nonumber \\
	& & \left( \frac{R}{D_1}  \cos{(\theta - \alpha_1)}  -  \frac{1}{[1 + (R/D_1)^2-2(R/D_1)\cos{(\theta-\alpha_1)}]^{1/2}}  \right).
\end{eqnarray}
Equation~\ref{eq:potfunc} is often expressed as an expansion in terms of the Legendre polynomials, $P_n$:
\begin{equation}\label{eq:legendre}
	V  =  \frac{G M_1}{D_1} \left[ 1+\sum_{n=2}^\infty \left(\frac{R}{D_1}\right)^n  P_n\left(\cos{(\theta-\alpha_1)}\right)\right].
\end{equation}
The tide-generating potential function, given by Eqs.~\ref{eq:potfunc} or \ref{eq:legendre}, can be used to obtain a formula for the height of the water above or below the equilibrium water height that would exist in the absence of tidal forcing due to a nearby gravitating body.  In particular, the difference in height can be computed from the tide-generating potential, $V$, the local acceleration of gravity, $g$, and a number of assumptions about the density and elasticity of the planet itself.

The density and elasticity of the planet are relevant because the planet is also deformed by the moon's gravity \citep{melchoir:1966fk}.  The overall effect of this deformation is to reduce the amplitude of the ocean tides by a factor of $(1 + k - h)$.  Here, $k$ is the love number representing the variation in the tide-generating potential produced by the planet due to the redistribution of the planet's material in response to the tide-generating potential of the moon(s).  $h$ is the love number representing the variation of the surface height of the planet.  Calculation of these love numbers relies upon the assumption that the planetary tides, like the ocean tides, are generated by the tide-generating potential, and that they can therefore be expanded in terms of Legendre polynomials. There are significant problems, however, with truncating the tide-generating potential, as we shall explain shortly.  In any case, in this paper we consider the case of a perfectly rigid planet, which implies that $h = k = l = 0$.  Consequently the tidal amplitude correction factor mentioned above is reduced to unity.  Thus the tidal height can be obtained by simply dividing the tide-generating potential by the local gravitational acceleration.  Since we are here interested only in classifying transitions between different tidal patterns, it is not critical at this point to obtain an exact expression for the tidal heights.  We henceforth simply identify the maxima and minima of the tide-generating potential function with the locations of the high and low tides, respectively, which are formed on the surface of the planet. 

It is sometimes convenient to transform the tide-generating potential from local coordinates ($\theta$ and $\alpha$) into geographic and astronomical coordinates using the formula
\begin{equation}\label{eq:transform}
\cos{(\theta - \alpha)} = \sin{(\phi)}\sin{(\delta)} + \cos{(\phi)}\cos{(\delta)}\cos{(H)}.
\end{equation}
Here, $\phi$ and $H$ are the geographic latitude and hour angle of the observer and $\delta$ is the declination of the moon \citep{Doodson:1921fk}.  Since this mathematical transformation can not change the physical structure of the predicted tidal pattern, and since it will increase the notational complexity of our discussion, we will not make use of Eq.~\ref{eq:transform}.

Returning to Eq.~\ref{eq:legendre}, when $R/D_1$ is small, the series converges quickly. This is the case for the earth-moon system, in which $R/D_1 \simeq 0.017$.  The $n=2$ term gives rise to two tidal maxima; the $n=3$ and higher terms deform these tides slightly, but are not large enough to produce additional maxima or minima in the potential function.  In this sense, truncating the expansion after the $n=2$ term will not change the predicted tidal pattern.   

On the other hand, when $R/D_1$ is sufficiently large, it might seem that the higher order terms in Eq.~\ref{eq:legendre} could give rise to not just two, but perhaps three or more tidal maxima.  A bit of reflection, however, reveals that this is not the case.  First, it should be noted that the Legendre polynomials form a complete set for a large class of functions, of which Eq.~\ref{eq:potfunc} is an example.  It would be incorrect to assume that a function that can be expanded in terms of Legendre polynomials must have multiple maxima simply because of the existence of high order terms in its expansion.  

\begin{figure}[tbp]
\begin{center}
\includegraphics[angle=0,width =5.5in]{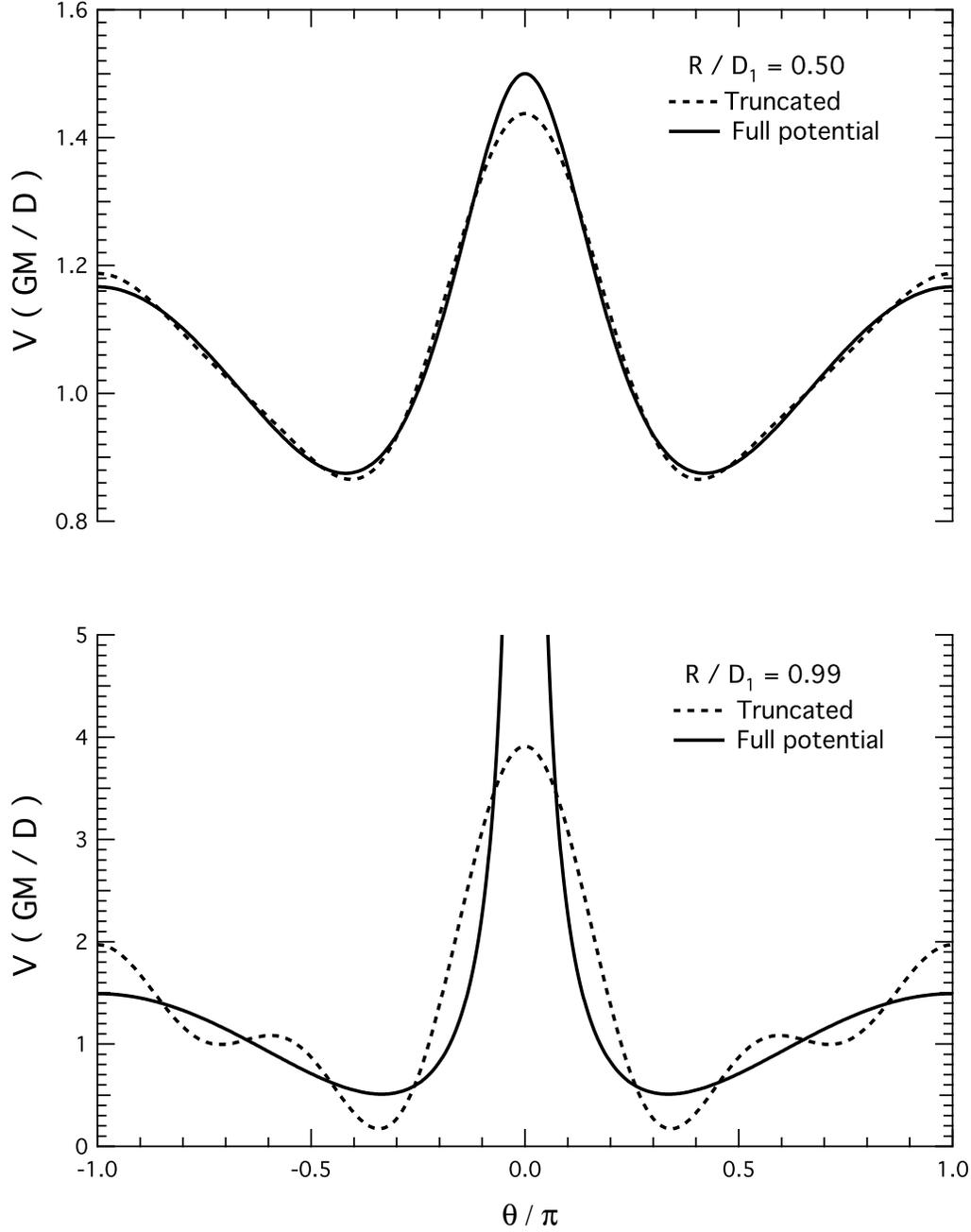}
\caption[fig2]{
\label{fig:fig2}
The tide-generating potential as a function of $\theta$ for a single moon orbiting a planet.  The solid line represents the closed form expression (Eq.~\ref{eq:potfunc}); the dashed line represent the series expansion (Eq.~\ref{eq:legendre}) truncated after the fourth order term.  The upper diagram shows that they agree when $R/D_1 = 0.5$; the lower diagram shows that they disagree when $R/D_1 = 0.99$.}
\end{center}
\end{figure}

It is true that Eq.~\ref{eq:legendre} could predict more than two maxima if one could independently vary the coefficients in front of the different order terms.  In the present case, however, the coefficients of the various $P_n$ are prescribed powers of $R/D_1$, and cannot be independently varied.  It is also true that Eq.~\ref{eq:legendre} could predict more than two maxima if one were to make the coefficient $R/D_1$ large, and then simply truncate the expansion after a few terms.  This, however, is not appropriate when $R/D_1$ is large, since in such a case the subsequent higher order terms are relevant in that they actually \textit{suppress} the appearance of more than two maxima in Eq.~\ref{eq:legendre}.  This important point is illustrated in Fig~\ref{fig:fig2}.  

This figure depicts the tide-generating potential as a function of $\theta$ for two values of $R/D_1$.  In both upper and lower diagrams, the solid line represents the closed form expression (Eq.~\ref{eq:potfunc}) and the dashed line represent the series expansion (Eq.~\ref{eq:legendre}) which has been truncated after the fourth order legendre polynomial.  The upper diagram shows that the two formulas predict the same number of high and low tides for the case of $R/D_1 = 0.5$.  The lower diagram shows that the two formulas predict a different number of high and low tides for the case of  $R/D_1 = 0.99$.\footnote{In the lower diagram it was necessary to limit the vertical range of the axis, and hence cut off the maximum of the closed form expression near $\theta = 0$, so as to illustrate the additional maxima and minima in the series expansion near $\theta/\pi = \pm 0.6$.  As $R/D_1$ approaches unity, the potential blows up at $\theta/\pi=0$.  This is an artifact of the unphysical scenario where the moon orbits \textit{on} the surface of the planet.  We chose the large value of $R/D_1$ in this diagram only to illustrate the different predictions of the closed form and the truncated series expansions of the tide-generating potential.}  Clearly this spurious behavior is an artifact of the truncation and is not a characteristic of the potential function itself.  For this reason, we henceforth do not resort to the legendre polynomial expansion of the potential.  Rather, we use the closed form expression for the potential, Eq.~\ref{eq:potfunc}.  This will allow us to altogether avoid truncation errors when dealing with arbitrarily large values of $R/D_1$.

Thus far we have limited our discussion to the tide-generating potential created by a single orbiting moon.  By the principle of superposition, it is clear that if $N$ moons orbit the planet, we can simply sum the tide-generating potential functions of the $N$ moons so as to achieve a generalized formula.
\begin{eqnarray}\label{eq:f5}
	V & = & - \sum_{i=1}^N\frac{G M_i  }{D_i} \times \nonumber \\
	& & \left(\frac{R}{D_i}  \cos{(\theta - \alpha_i)}   - \frac{1} { [ 1+(R/D_i)^2-2 (R/D_i)\cos{(\theta-\alpha_i)}]^{1/2}  } \right)
\end{eqnarray}
Here, $M_i$, $D_i$, and $\alpha_i$ are the lunar mass, lunar distance, and angular position of the $i^{th}$ moon.  Eq.~\ref{eq:f5} is the fundamental formula from which can be found the locations of the high and low tides at any instant at polar coordinates $(R,\theta)$ in the lunar orbital plane of a planet orbited by $N$ moons.  In the remainder of the paper, we analyze the properties of Eq.~\ref{eq:f5}.

\section{Tidal action for two moons}
\label{sec:2moons}

\subsection{Identical lunar masses and distances}
\label{subsec:samemassesdistances}

Let us restrict our attention at first to the case of two moons orbiting a planet.  It will prove convenient henceforth to refer to $M_2/M_1$ as the mass ratio, and to $D_2/D_1$ as the distance ratio.  Let us consider the special case in which both the mass ratio and the distance ratio are unity:  $M_1=M_2\equiv M$ and $D_1=D_2\equiv D$.  For symmetry, we orient the coordinate system such that $\alpha_1 = -\alpha_2 \equiv \alpha$.  The parameter $\alpha$ thus describes the lunar separation half-angle.  Eq.~\ref{eq:f5} now becomes
\begin{eqnarray}\label{eq:f6}
	V & = &  \left( \frac{1}{[1+(R/D)^2-2(R/D)\cos{(\theta-\alpha)}]^{1/2}} \right)\nonumber \\
	& + &   \left( \frac{1}{[1+(R/D)^2-2(R/D)\cos{(\theta+\alpha)}]^{1/2}} \right) \nonumber \\
	& - &   \left(\frac{2R}{D}\cos{(\theta)}\cos{(\alpha)} \right).
\end{eqnarray}
We shall henceforth refer to the quantity $R/D$ as the planetary radius\footnote{Note that increasing the planetary radius is equivalent to decreasing the lunar orbital distance.} and the quantity $\alpha/\pi$ as the lunar angle.  In Eq.~\ref{eq:f6}, we have normalized so that the potential is measured in units of $GM/D$.
 
\begin{figure}[tbp]
\begin{center}
\includegraphics[angle=0,width =5.5in]{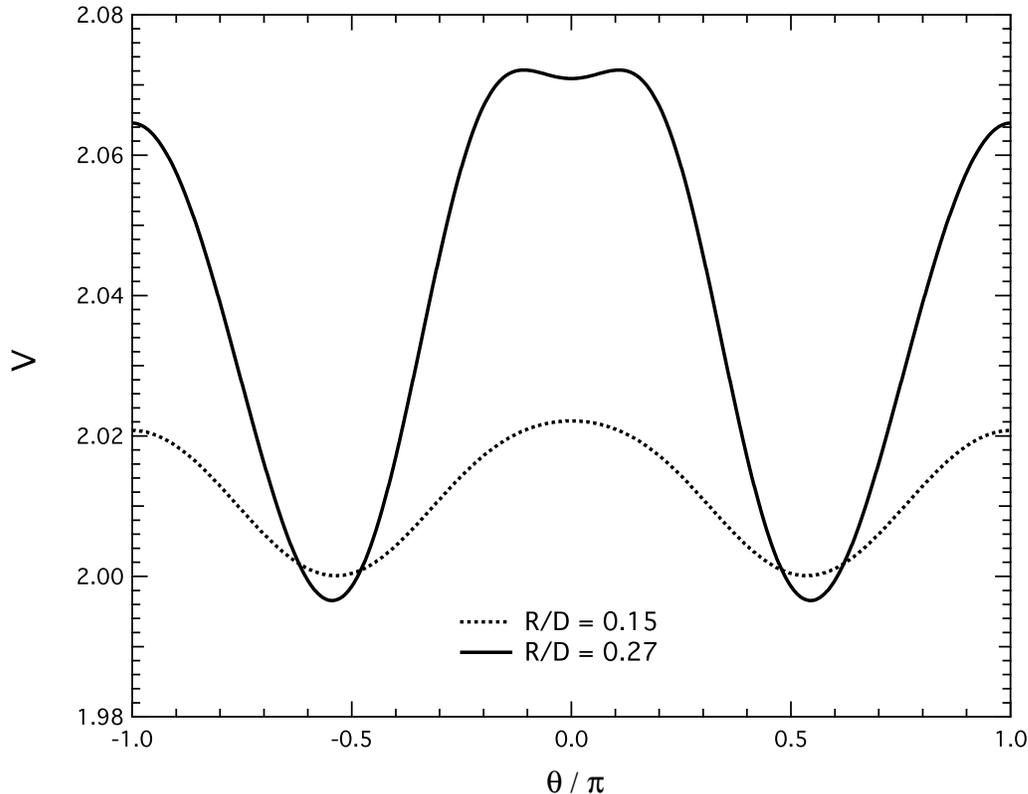}
\caption[fig3]{
\label{fig:fig3}
The tide-generating potential as a function of $\theta$ for $R/R_c<1$ (dotted line) and $R/R_c>1$ (solid line).  The potential is measured in units of $Gm/D$.  The instantaneous angular position of the moons are here taken to be $\alpha / \pi = \pm 0.2$. }
\end{center}
\end{figure}

Eq.~\ref{eq:f6} is plotted as a function of $\theta$ for two values of planetary radius in Fig.~\ref{fig:fig3}.  In this diagram, we have chosen to use a lunar angle of $0.2$.  First, consider the dotted line, representing the case in which the planetary radius is $0.15$.  There exist high tides at $ \theta / \pi = 0 \mbox{ and } 1$, and low tides at $ \theta / \pi \approx \pm 0.52$.  This is consistent with a well known tidal pattern, an ellipse, with its major axis aligned along the abscissa of our coordinate system.  

Now consider the solid line in Fig.~\ref{fig:fig3}, representing the case in which the planetary radius is $0.27$.  The high tide at $\theta / \pi = 1$ is still at $\theta / \pi = 1$.  However, now the point at $\theta = 0$ has become a low tide.   The high tide at $\theta = 0$, when the planetary radius was $0.15$, has split into two high tides separated by a low tide at $\theta = 0$.  A bifurcation has occurred at some intermediate value of the planetary radius.  Previously, the second order multipole moment dominated the tide generating potential.  The bifurcation signals the point at which higher order multipole moments become significant.  As a result of the bifurcation, the tidal pattern no longer resembles an ellipse, but rather has three lobes, or three high tides.  The angular orientation of these lobes will be discussed shortly.  Generally speaking, the number of high and low tides is indicated by the number of maxima and minima occurring in Eq.~\ref{eq:f6} for a given value of the planetary radius.   It should be noted that high tides are always separated by low tides, so the number of high and low tides are always equal.  

\begin{figure}[tbp]
\begin{center}
\includegraphics[angle=0,width =5.5in]{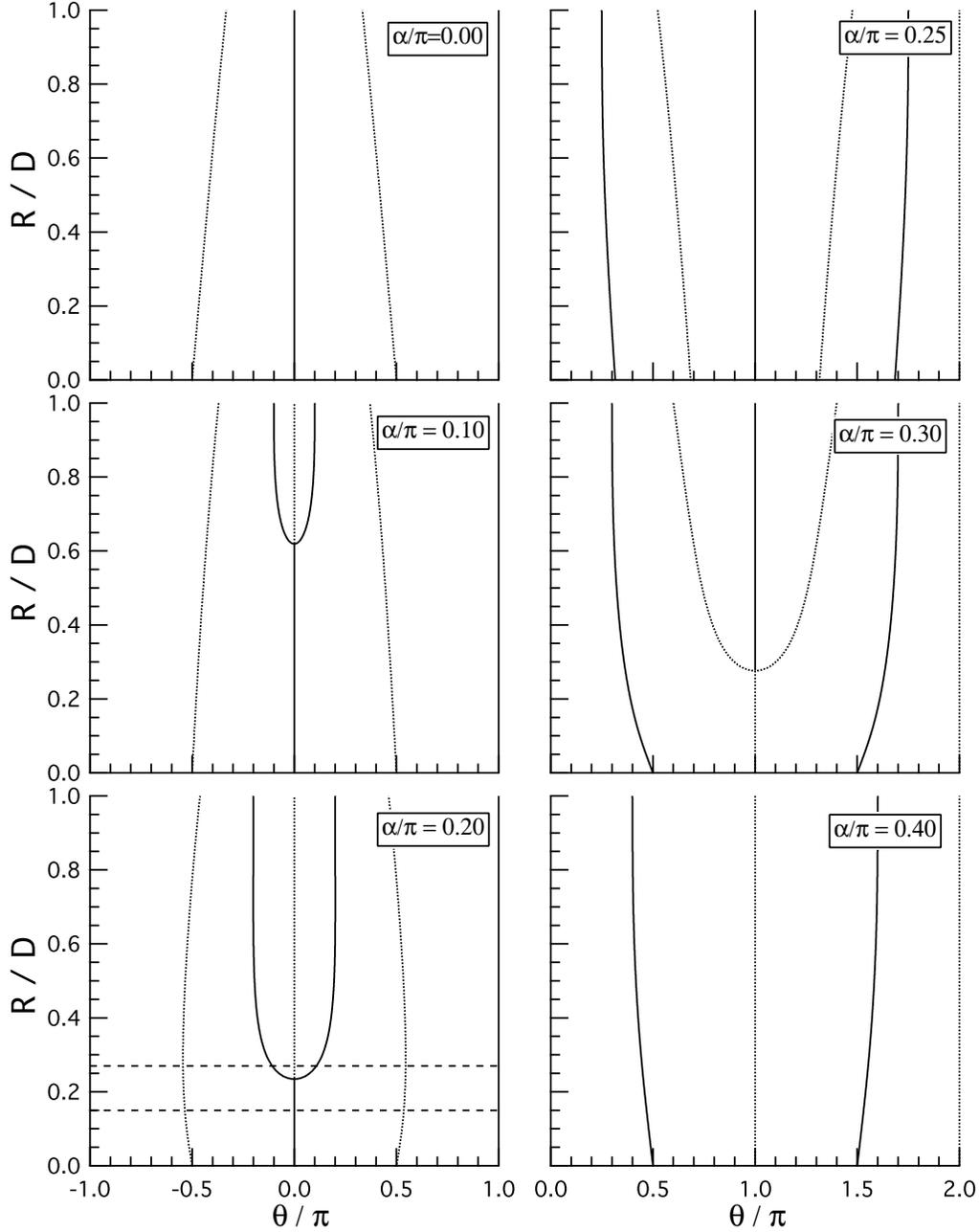}
\caption[fig4]{
\label{fig:fig4}
Bifurcation plots for various values of the lunar angle, $\alpha/\pi$.  The angular position, $\theta$, of the high (low) tides are indicated by solid (dotted) lines.
}
\end{center}
\end{figure}

In order to determine the critical value of the planetary radius at which the bifurcation occurs between the two tidal patterns for a given lunar angle, we can make a plot of the maxima and minima of Eq.~\ref{eq:f6} as a function of both planetary radius and $\theta$.  Such plots are shown in Fig.~\ref{fig:fig4} for six different values of the lunar angle.  Solid lines represent the angular positions of high tides; dotted lines represent the angular positions of low tides.  Note that the right edge of each plot has either a solid or a dotted line, indicating a particular type of tide.   

Several important points should be made regarding Fig.~\ref{fig:fig4}.  Recall that Fig.~\ref{fig:fig3} depicted the tide-generating potential for a lunar angle $\alpha = 0.20$ for the special cases of $R/D = 0.15 \mbox{ and } 0.27$.  Consider now the plot in the lower left corner of Fig.~\ref{fig:fig4}.  This is the bifurcation plot for the same lunar angle.   The values $R/D = 0.15 \mbox{ and } 0.27$ are indicated by horizontal dashed lines.  Notice that when $R/D =0.15$, there are two high tides ($\theta /\pi = 0$ and $\theta / \pi =1$) and two low tides ($\theta/\pi \approx \pm 0.52$).  Most notably, a bifurcation occurs near $R/D \approx 0.23$ where the high tide at $\theta / \pi =0$ splits into two high tides.  Therefore when $R/D =0.27$, there are three high tides ($\theta /\pi \approx \pm 0.12$ and $\theta / \pi =1$) and three low tides ($\theta/\pi \approx \pm0.55$ and $\theta / \pi = 0$).  Generally speaking, for lunar angles between zero and $0.25$, the high tide at $\theta =0$ bifurcates into two high tides at a critical planetary radius,  $R_c/D$, which depends on the lunar angle.  When $R/D > R_c/D$, the angular positions of the newly formed high tides vary with planetary radius, approaching the angular positions of the two moons as the planetary radius approaches unity.  In other words, when the moons are very close to the surface of the planet, the high tides tend to be directly beneath the moons.  

Now consider the plot in the upper left corner.  When the two moons are very close to one another, the lunar angle approaches zero.  In the limit in which the lunar angle is zero, there are just two low tides and two high tides.  There is no bifurcation.  In this case, the angular positions of the low tides depend on the planetary radius, but those of the high tides are constant, at $\theta /\pi = 0 \mbox{ and } 1$.  Although it is impossible for two moons to be on top of one another, the case of zero lunar angle can be interpreted physically as the case of just a single moon orbiting a planet.  This demonstrates that, within the equilibrium conditions assumed here, it is not possible for more than two high tides to form when a single moon orbits a planet, as was stated in Sec.~\ref{sec:bg}.
 
Finally, consider the plots in the right hand column.  For a lunar angle of $0.25$ (when the moons are in quadrature) there exist three high tides for all values of the planetary radius.   For lunar angles greater than $0.25$, it is not the high tide at $\theta = 0$ that bifurcates into two high tides, but rather the low tide at $\theta / \pi= 1$ that bifurcates into two low tides, which straddle a newly formed high tide at $\theta / \pi =1$.  Again, as the planetary radius approaches unity, the angular positions of two of the high tides approach the angular positions of the two moons.  The angular position of the third high tide is always at $\theta / \pi =1$.  

The results presented here will be revisited in the stability diagram presented in Sec.~\ref{subsec:diffdistances}, where we consider the more general case in which the moons are not constrained to orbit at the same distance.   

\subsection{Identical lunar masses, different distances}
\label{subsec:diffdistances}

We retain the assumption that the mass ratio is unity.  We use the same procedure as before, except that in Eq.~\ref{eq:f6} we fix $D_1$ and allow both the planetary radius, $R/D_1$, and the distance ratio, $D_2/D_1$, to vary.  Since the orbital distances differ, the lunar angle will be changing over time as the moons orbit the planet at different angular speeds.  In our analysis we are assuming this is happening sufficiently slowly on the time scale of the fluid response that the equilibrium theory is justified.  

\begin{figure}[tbp]
\begin{center}
\includegraphics[angle=0,width =5.5in]{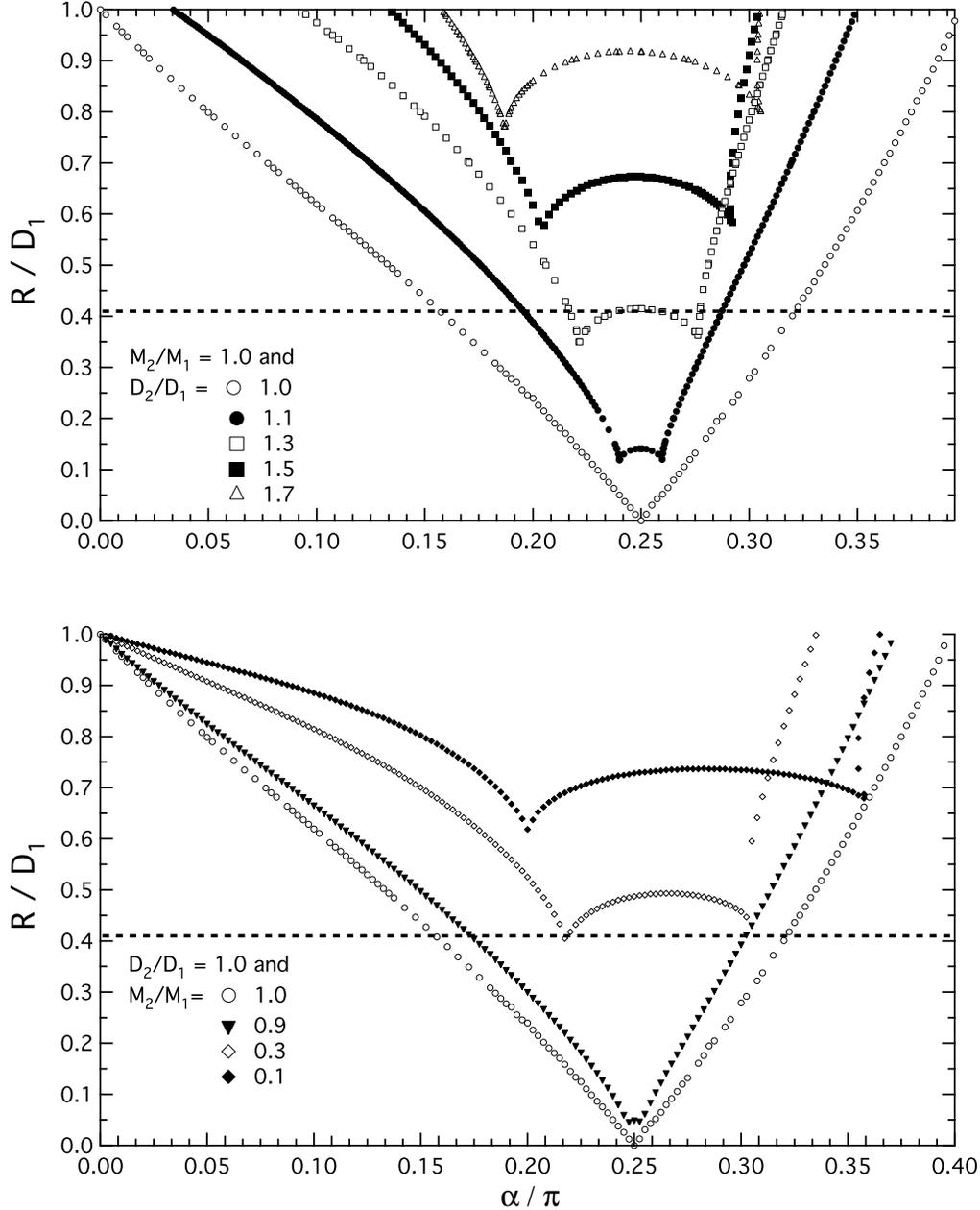}
\caption[fig5]{
\label{fig:fig5}
Stability boundaries for the case of two moons orbiting a planet of radius $R$.  There are three (two) high tides in the regions above (below) each stability boundary.  Top:  Constant mass ratio; varying distance ratio. Bottom:  Constant distance ratio; varying mass ratio.  The horizontal dashed line represents the Roche limit.}
\end{center}
\end{figure}

The bifurcation plots for this scenario (not shown) are similar to the ones depicted in Fig.~\ref{fig:fig4} in that they reveal a critical planetary radius, $R_c/D_1$, at which a transition between two and three high tides occurs.  However, the value of the critical planetary radius now depends upon both the lunar angle and the distance ratio.  This is illustrated in the stability diagram of Fig.~\ref{fig:fig5}.  

The top diagram of Fig.~\ref{fig:fig5} depicts the case in which the mass ratio is unity and the distance ratio is permitted to vary.  The legend indicates which symbols correspond to which value of the distance ratio.  For clarity, we only depict data for five values of the distance ratio.  As an example of how to read the top diagram of Fig.~\ref{fig:fig5}, consider the open circles which appear in that plot.  These correspond to the special case, considered in Sec.~\ref{subsec:samemassesdistances}, in which the two moons orbit at the same distance from the planet.  From the bifurcation plots of Fig.~\ref{fig:fig4}, recall that for $\alpha = 0.20$, a bifurcation occurs when $R/D_1 \approx 0.23$.  Correspondingly, there is an open circle in the top diagram of Fig.~\ref{fig:fig5} at the coordinates ($\alpha/ \pi = 0.20$, $R/D_1 \approx 0.23$).  The string of open circles, then, represents a stability boundary.  Above the stability boundary, three high tides are stable; below the stability boundary, two high tides are stable.  

We can make the following general observations regarding the top diagram of Fig.~\ref{fig:fig5}.  First, the location of the stability boundary depends upon the distance ratio of the moons.  As the distance ratio increases from $1.0$ toward $2.0$, the region in which three high tides are stable becomes narrower.  

Physically, this occurs because when the second moon is farther from the planet's surface, its gravitational field is less capable of raising a separate tide.  Not shown in the figure is that when the mass ratio is unity and the distance ratio exceeds 1.94, the region of three high tides vanishes entirely, leaving only two high tides.  

Second, when the distance ratio is unity, the critical planetary radius has a value of zero at lunar angle $0.25$.  This point we henceforth refer to as a \textit{zero (radius) stability point}.  This means that at this lunar angle, three high tides will be stable for any planetary radius.  We will discuss the zero stability point in greater detail in Sec.~\ref{subsec:diffmassesdistances}.

Third, when the distance ratio differs from unity, there is no longer a zero stability point, and the critical planetary radius generally has minima at two different lunar angles.  This implies that for such a system, as the lunar angle changes during the course of the lunar orbits, the planet can experience alternating periods of high and low tides. Consider, for instance, a hypothetical system for which $R/D_1 \approx 0.41$ (the horizontal dashed line) and for which $D_2/D_1 = 1.3$.  As the lunar angle changes between zero and $0.40$, the planet will experience first two, then three, then two, then three, and finally two high tides.  

\subsection{Identical distances, different masses}
\label{subsec:diffmasses}

Now let us fix the distance ratio at unity and allow the mass ratio to vary.  This is depicted in the bottom diagram of Fig.~\ref{fig:fig5}.  Like the top diagram, it depicts the stability boundary between regions of parameter space in which two and three high tides are stable.  Unlike the top diagram, the various symbols now represent different values of the mass ratio, as noted in the legend.  For clarity, we only show data for four values of the mass ratio between 0.1 and 1.0.  It is apparent from this diagram that the instability persists even when the masses are not identical.  Qualitatively, as far as the shape of the stability boundaries is concerned, the effect of reducing the mass of the second moon is similar to that of increasing its distance from the surface of the planet.  In either case, the second moon is less capable of raising a separate tide. 

\subsection{The Roche Limit}
\label{subsec:roche}

In the previous two sections, we considered separately the effects of lunar mass and lunar orbital radii on the formation of tidal patterns.  In the following section, we will combine these results so as to derive a formula which prescribes the conditions which would render the formation of more than two high tides likely.  Before proceeding, however, we should ask whether the orbital radii which we have been considering are in fact physically obtainable.  That is: is it reasonable to suggest that the planetary radius can range anywhere between almost zero and almost unity?

Obviously, when the planetary radius is close to unity, the moon is orbiting practically \textit{on} the surface of the planet, which is unrealistic.  But what is a reasonable value of $R/D$ for a lunar orbit?  The Roche limit \citep{Roche:1847fk} is the planetary radius above which the host planet's tidal forces acting on the orbiting moon exceed the moon's gravitational self-attraction.  If the planetary radius is larger than the Roche limit,  an orbiting moon would probably not have formed by accretion.  The exact value of the Roche limit depends upon the rigidity and density of the moon.  Assuming a fluid moon whose density is the same as that of the host planet, the Roche limit is approximately $0.41$; a rigid moon yields a value closer to $0.79.$  The Roche limit of $0.41$ is depicted in both the upper and lower diagrams of Fig.~\ref{fig:fig5} as a horizontal dashed line.  It is notable that there still exists a significant region of parameter space ($0<R/D<0.41$) such that a moon formed by accretion could generate multiple tides on the surface of its host planet.  The less conservative value of the Roche limit yields an even larger region.

\subsection{Different distances and masses}
\label{subsec:diffmassesdistances}

\begin{figure}[tbp]
\begin{center}
\includegraphics[angle=0,width =5.5in]{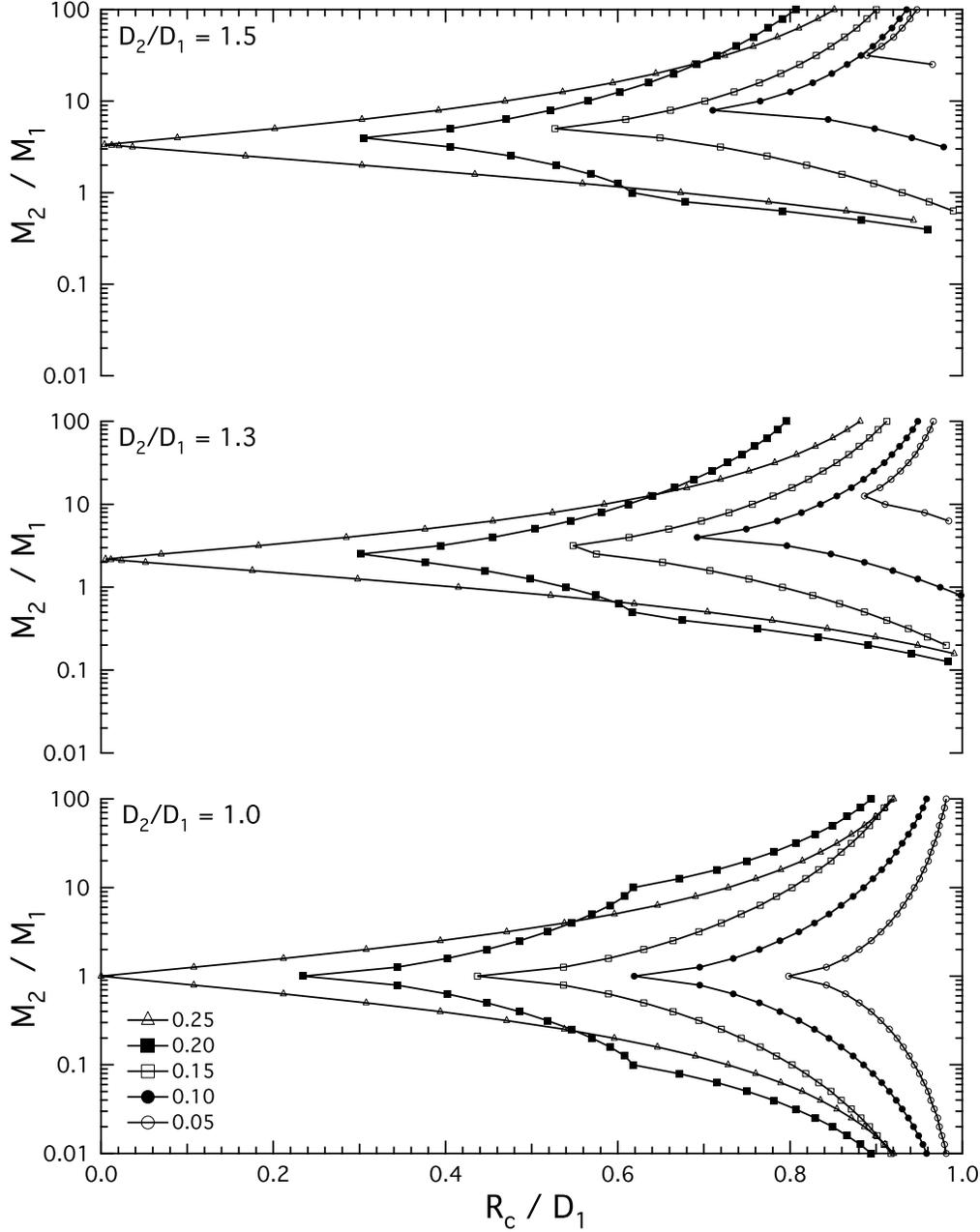}
\caption[fig6]{
\label{fig:fig6}
Variation of the stability boundary with mass ratio.   Values of the lunar angle are shown in the legend.  From bottom to top, the distance ratios are 1.0, 1.3, and 1.5, respectively.   Notice that the zero stability point, located at $R_c/D_1=0$, moves to higher mass ratio as the distance ratio increases.}
\end{center}
\end{figure}

Now let us consider scenarios in which both the distance and the mass ratios differ from unity.  In Fig.~\ref{fig:fig6}, we show how the stability boundaries vary over four decades of the mass ratio, for three different values of the distance ratio.  The stability diagrams in Fig.~\ref{fig:fig6} are similar to those in Fig.~\ref{fig:fig5}, except that they are turned on their side, and that the planetary radius is plotted versus the mass ratio instead of the lunar angle.  Fig.~\ref{fig:fig6} is designed to highlight the variation of the zero stability point, defined in Sec.~\ref{subsec:diffdistances}, with mass ratio.

In the bottom diagram, the distance ratio is fixed at unity.  The  legend indicates five values of the lunar angle between 0.05 and 0.25.  For instance, at lunar angle $0.25$ (denoted by triangles), the critical planetary radius is zero when the mass ratio is unity.   This is our zero stability point.

In the middle and top diagrams of Fig.~\ref{fig:fig6}, we show how the stability boundaries vary with mass ratio when the distance ratios are 1.3 and 1.5, respectively.    Now the zero stability point occurs when the mass ratio is larger then unity.  Notice that the zero stability point always occurs at a lunar angle of $0.25$.

\begin{figure}[tbp]
\begin{center}
\includegraphics[angle=0,width =5.5in]{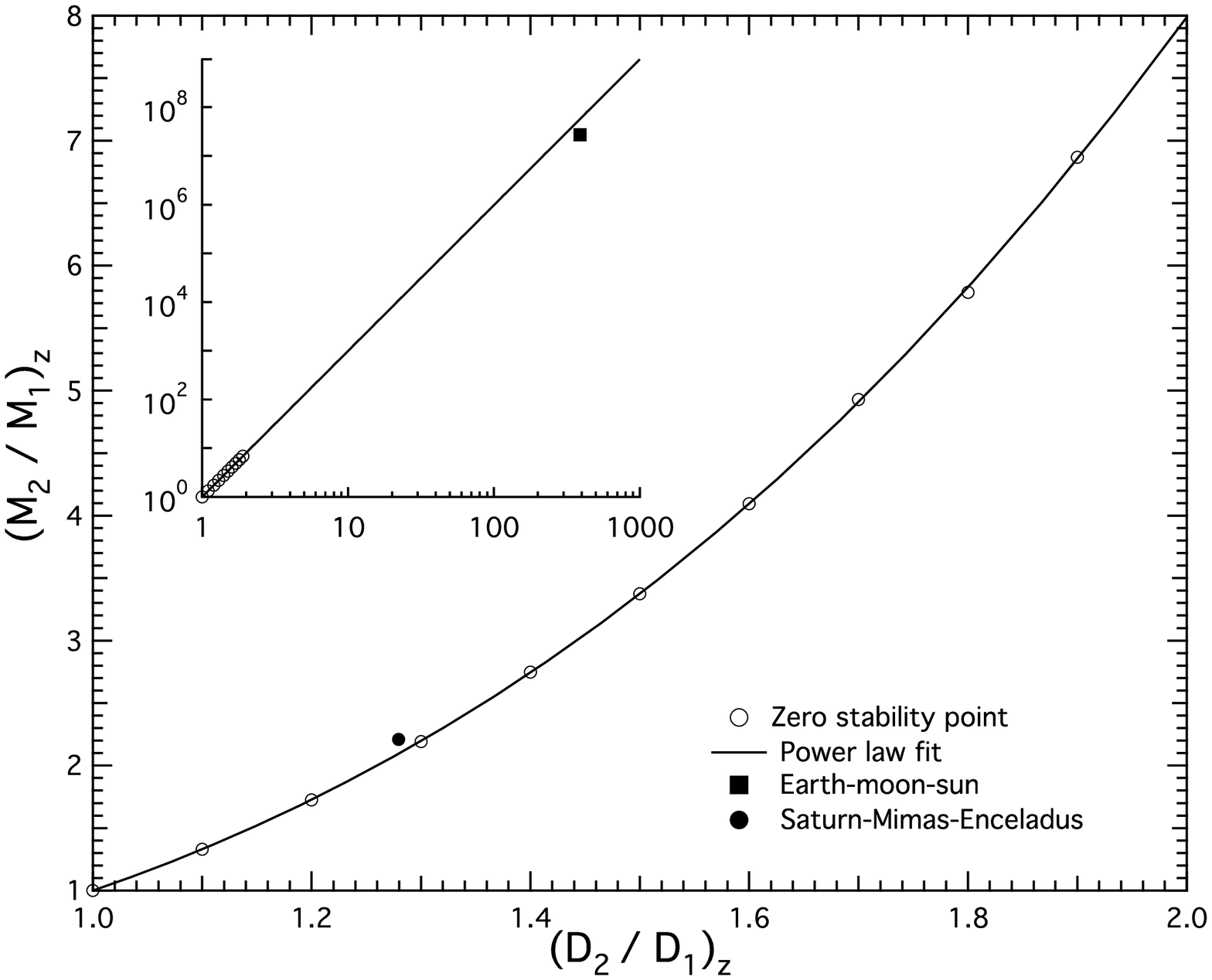}
\caption[fig7]{
\label{fig:fig7}
Variation of the calculated zero stability point with distance ratio.  The fit is a power law with exponent 3.  The inset depicts an extrapolation of the power law fit over three decades of distance ratio.  The closed circle represents the relationship between the mass ratio and the distance ratio for Mimas and Enceladus, two moons of Saturn.   The closed square represents the relationship between the mass ratio and the distance ratio for the moon and the sun relative to the earth.}
\end{center}
\end{figure}

One can study how the zero stability point depends upon the distance ratio.  This is depicted in Fig.~\ref{fig:fig7}.  The open circles represent the calculated zero stability points.  The solid circle represents the lunar mass and distance ratios of the Saturnian moons Mimas and Enceladus (we shall return to this shortly).  The solid line in Fig.~\ref{fig:fig7} is a power law fit, 
\begin{equation}\label{eq:f7}
( M_2/ M_1)_z = A + B\  ( D_2 / D_1)_z^{\gamma},
\end{equation}
to the calculated zero stability points.  The subscript, $z$, indicates that these are the lunar mass and distance ratios at the zero stability point.  The fitting parameters have the values $A = 0.00 \pm 0.04$, $B = 1.00 \pm 0.03$ and $\gamma = 2.99 \pm 0.04$.  The uncertainties in the fitting parameters indicate one standard deviation confidence intervals.  

Our analysis implies that for any value of the distance ratio, there is a corresponding value of the mass ratio at which there must be three high tides, when the moons are in quadrature.  Moreover, our analysis implies that when Eq.~\ref{eq:f7} only approximately holds, \textit{there still exists a significant range of lunar angles over which three high tides will form for a particular planetary radius}.  Practically, Eq.~\ref{eq:f7} serves as a guide in predicting which two-moon systems are likely to exhibit three high tides.  Physically, Eq.~\ref{eq:f7} states that there can be three high tides on a planet with one very distant moon and one very near moon, provided that the more distant moon is more massive than the nearby moon.  Recall, however, that moons with different orbital radii will have different orbital periods, and hence the lunar angle will not be constant.  Consequently, a two moon planet could experience three high tides for some duration, and two high tides for a different duration, of the time it takes the moons to orbit the planet. 

\subsection{Earth-moon-sun, Saturn-Mimas-Enceladus, and other systems}
\label{subsec:ems}

Let us use the equilibrium analysis developed thus far to predict the tidal pattern on the surface of the earth from the combined action of the sun and the moon.  The solid square in the inset of Fig.~\ref{fig:fig7} represents the mass and distance ratios for the moon, $M_1$, and the sun, $M_2$, orbiting the earth.\footnote{Practically, we can consider the sun to act like a second moon orbiting the origin of earth's non-inertial reference frame, for purposes of calculating earth's ocean tides.  There are complications, which we are neglecting, that arise from the fact that the orbital plane of the moon is oblique to the ecliptic.}  These values are provided in Tab.~\ref{tab:tab1}.  The solid line in the inset of Fig.~\ref{fig:fig7} is the power law fit to our calculated zero stability points extrapolated over approximately three decades of mass ratio.  Incidentally, the earth-moon-sun system lies fairly close to the extrapolated line of zero stability points.  This seems to suggest that there might be three high tides on the earth when the moon and sun are in quadrature.  We show next that this is not, however, the case.

\begin{figure}[tbp]
\begin{center}
\includegraphics[angle=0,width =5.5in]{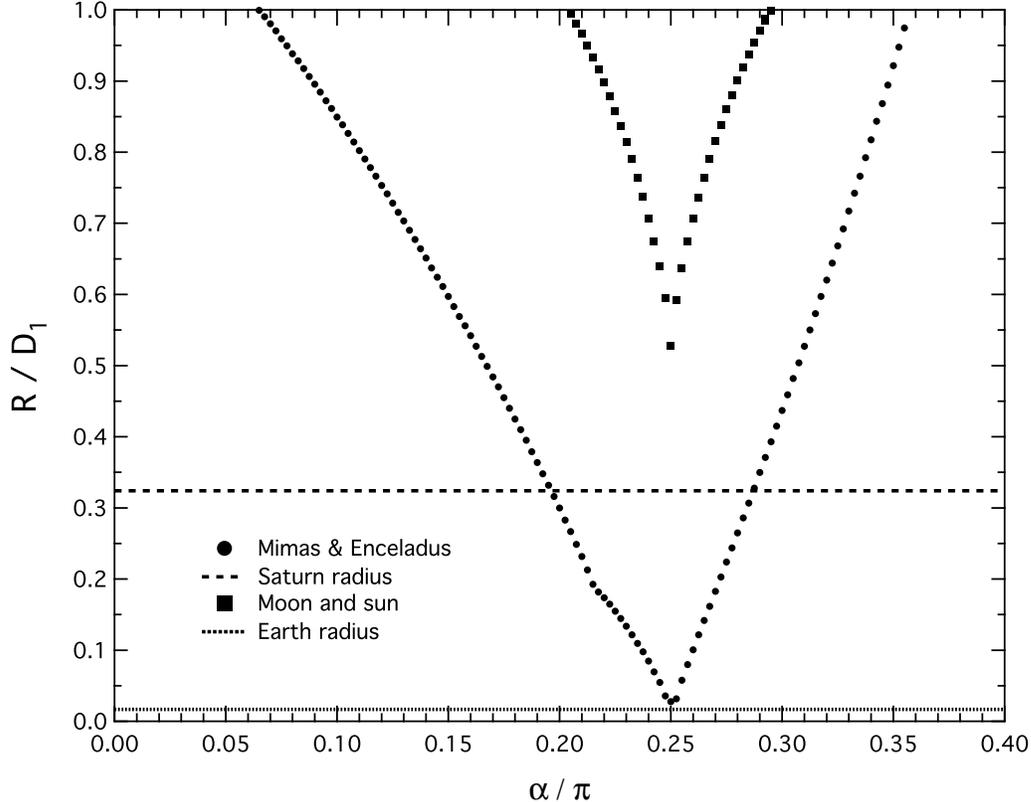}
\caption[fig8]{
\label{fig:fig8}
Stability diagram depicting the stability boundary for the case of Mimas and Enceladus orbiting Saturn (closed circles) and the moon and the sun orbiting the earth (closed squares).  The horizontal dashed line at $R/D_1 = 0.324$ represents the planetary radius of Saturn.  The horizontal dotted line at $R/D_1 = 0.017$ represents the planetary radius of the earth.}
\end{center}
\end{figure}

In Fig.~\ref{fig:fig8} we show a stability diagram, much like those in Fig.~\ref{fig:fig5}.  In this case, however, the closed squares form the stability boundary specifically for the earth-moon-sun system.  The reason that the earth exhibits two, rather than three, high tides each day is that the planetary radius of the earth is approximately $R/D = 0.017$.  This planetary radius is depicted as a horizontal dotted line near the bottom of the plot.  Clearly this lies far below even the lowest point on the stability boundary.  Our analysis is therefore consistent with the existence of semi-diurnal tides on the earth.  Were the earth larger, such that $R/D_1  \approx 0.55$, then we could expect three high tides, rather than two high tides, during certain phases of the moon.  Interestingly, the moon was likely much nearer the earth shortly after its origin, implying that such conditions may have been fulfilled early in the moon's history \citep{Hartmann:1984uq}.

The earth-moon-sun system does not exhibit three high tides.  Should we expect any realistic system to exhibit three high tides?  Having two moons, Mars is a candidate.  Unfortunately, Mars has no oceans.  Furthermore, Mars' more distant moon, Deimos, is less massive than its nearby moon, Phobos.  Therefore even if Mars did have surface oceans, we would expect it to have only semi-diurnal tides.  Considering non-terrestrial planets, both Saturn and Jupiter have many moons, but no surface oceans.  It just so happens, however, that two of Saturn's moons, Mimas and Enceladus, satisfy the conditions to produce three high tides on the surface of Saturn:  the Saturn-Mimas-Enceladus system lies very close to the line of predicted zero-stability points in Fig.~\ref{fig:fig7}.  For the sake of argument, let us suppose that these were Saturn's only two moons.  In Fig.~\ref{fig:fig8}, the closed circles form the stability boundary for the Saturn-Mimas-Enceladus system.  Notice that the stability boundary dips to very low values of the planetary radius.  The horizontal dashed line at $R/D = 0.324$ represents the planetary radius of Saturn.  If Saturn only had these two moons, one would expect three high tides on the surface of Saturn for lunar angles between approximately $0.20$ and $0.28$.  

On the other hand, if Io and Europa were Jupiter's only two moons, they would not raise three high tides on the surface of Jupiter.  This is because the more distant moon, Europa, is less massive than the nearby moon, Io.

\section{Three moons}
\label{sec:3moons}

\begin{figure}[tbp]
\begin{center}
\includegraphics[angle=0,width =5.5in]{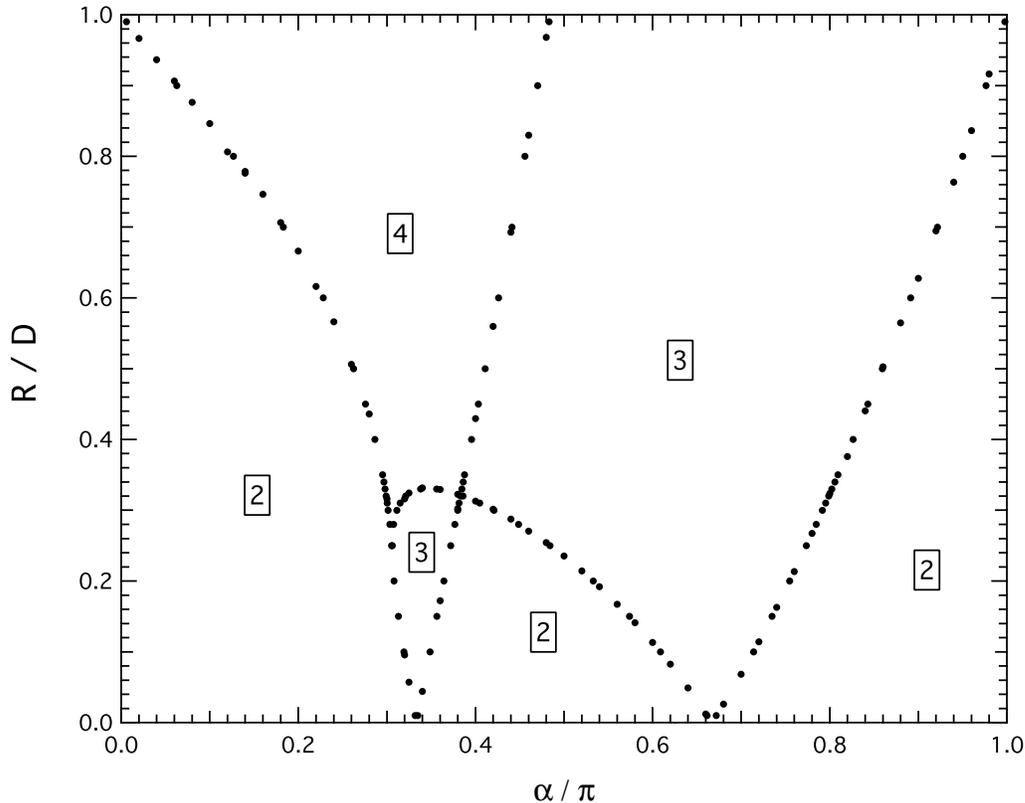}
\caption[fig9]{
\label{fig:fig9}
Stability diagram for three identical moons orbiting a planet, radius $R$, at identical radii, $D$.  The angular locations of the moons are at $0$ and $\pm \alpha$.  The planetary radius is shown on the ordinate, the lunar angle on the abscissa.  The insets describe the number of high tides occurring in each region of the diagram.}
\end{center}
\end{figure}

In principle, the tidal structure of any $N$ moon planet can be determined from Eq.~\ref{eq:f5}.  As a (relatively) simple example, we consider briefly the case of a three-moon planet.  We assume that the moons have identical masses and orbital distances, and we place them at lunar angles $0$ and  $\pm \alpha$.  The stability diagram for this system is illustrated in Fig.~\ref{fig:fig9}.   It displays a somewhat more complicated structure.  The number of high tides in each region of parameter space is indicated by a boxed integer in that region.  It should be noted that any number of high tides between two and four can be stable in some region of parameter space.   In general, we suspect that for $N$ moons, any number of high tides between 2 and $N+1$ could, in principle, be stable.

\section{Conclusion}
\label{sec:conclusion}

We have studied the tide-generating forces which act on the surface of a planet covered by a uniform layer of fluid and orbited by multiple moons.  We found that the tidal pattern depends upon the orbital distances and the angular separation of the moons.  In particular, we found a tidal instability at a critical radius, $R_c/D$, which depends upon the lunar angles, the mass ratios, and the distance ratios of the orbiting moons.  The tidal instability gives rise to transitions between states in which different numbers of high tides are stable.  Physically, the tidal instability can be thought of as a transition between states in which different multipole moments of the tide-generating potential are dominant.

We have compared our results to those of \citet{Kapoulitsas:1985fk}, who, in the spirit of Newton's equilibrium theory, derived a formula for the height of the water surface on a planet orbited by two gravitating bodies, which he considered to be the sun and the moon.  After computing the partial tides due to both the sun and the moon, he concluded that, ``to the order of approximation adopted, the tidal height of the combined action of [the moon and the sun] is obtained by adding the tidal heights due to each."   

Our results agree with those of Kapoulitsas in the limiting case that there are two gravitating bodies and that these are taken to be the sun and the moon, with the appropriate mass and distance ratios.  In the general case, however, our results disagree with those of Kapoulitsas.  He arrives at his formulae for the tidal height by truncating the expansion of the tide-generating potential function in powers of the planetary radius, keeping only terms up to second order.  He does this so as to arrive at an analytic expression for the tidal height, which in turn supports his assertion that the tidal height caused by the combined action of the two gravitating bodies is approximated by adding the tidal heights due to each independently.  His final formula, Eq. 3.5, however,  implies that two identical moons orbiting at the same distances would never produce three high tides in the lunar orbital plane.  Rather, they would produce two high tides in every case except when the moons are in quadrature, in which case there would be no tidal height variation at all.  

This is in marked contrast with the results presented in this paper.  In light of the results presented here, this problem arises because the higher order terms in the tide-generating potential function are omitted.  We have demonstrated that in order to correctly identify the nature of the transition between different tidal patterns on multi-moon planets, it is necessary to utilize a closed-form expression for the tide-generating potential function.  This provides a technique for predicting tidal phenomena on putative extra-solar planets orbited by multiple moons.  It is conceivable that an extra-solar planet might fulfill the conditions described in this paper which would allow for the stability of three or more high tides over some region of time during the lunar orbits.  Generally speaking, the study of tidal phenomena on extra-solar planets could prove extremely useful in exploring their composition and formation.

\appendix




\bibliography{}

\bibliographystyle{elsart-harv}

\label{lastpage}

\clearpage


\begin{table}
\begin{center}
\begin{tabular}{llllllll}
Planet&Satellite &R (km)&M (kg)&D (km)&$R/D_1$&$D_2/D_1$&$M_2/M_1$ \\ \hline \hline \\
Earth&&$6.38\times10^{3}$&$5.97\times10^{24}$&&0.017&389&$2.72\times10^{7}$ \\
&Moon& &$7.36\times10^{22}$&$3.84\times10^{5}$&&& \\
&Sun&&$1.99\times10^{30}$&$1.50\times10^{8}$&&& \\ \hline \\
Mars&&$3.40\times10^{3}$&$6.42\times10^{23}$&&0.377&2.56&0.167 \\ 
&Phobos& &$1.08\times10^{16}$&$9.00\times10^{3}$&&& \\
&Deimos& &$1.80\times10^{15}$&$2.30\times10^{4}$&&& \\ \hline \\
Saturn&&$6.03\times10^{4}$&$5.68\times10^{26}$&&0.324&1.28&2.21 \\ 
&Mimas& &$3.80\times10^{19}$&$1.86\times10^{5}$&&& \\
&Enceladus& &$8.40\times10^{19}$&$2.38\times10^{5}$&&& \\ \hline \\
\end{tabular}
\end{center}
\caption[tab1]{
\label{tab:tab1}
Some, or all, of the satellites orbiting each planet are listed in the second column.  $R$ is the radius of the planet.  $M$ is the mass of the satellite or the planet.  $D$ is the distance of the satellite from the planet.  $R/D_1$ is the radius of the planet divided by the distance to the nearer of the two satellites.  $D_2/D_1$ is the distance ratio of the two satellites.  $M_2/M_1$ is the mass ratio of the two satellites. 
}
\end{table}

\clearpage	

\end{document}